\documentclass[journal]{IEEEtai}
\usepackage[algo2e,ruled,vlined,linesnumbered]{algorithm2e}
\usepackage[colorlinks,urlcolor=blue,linkcolor=blue,citecolor=blue]{hyperref}
\usepackage{mathtools}
\usepackage{color,array}
\usepackage{amsmath,latexsym,amssymb,amsfonts}
\usepackage{tabularx}
\usepackage{graphicx}
\usepackage{listings}
\usepackage{booktabs, makecell, tabularx}
\usepackage{subfig}
\usepackage{comment}
\usepackage{multirow,booktabs}

\begin{document}

\title{Variational energy based XPINNs for phase field analysis in brittle fracture} 

\author{Ayan Chakraborty,Cosmin Anitescu, Somdatta Goswami, Xiaoying Zhuang and Timon Rabczuk

\thanks{Ayan Chakraborty is with the Institute of Photonics, Leibniz University, Hannover, Germany (e-mail: ayan.chakraborty@iop.uni-hannover.de).}
\thanks{Cosmin Anitescu is with the Institute of Structural Mechanics, Bauhaus University-Weimar, Germany. (e-mail: cosmin.anitescu@uni-weimar.de).}
\thanks{Somdatta Goswami is with the Division of Applied Mathematics, Brown University, USA. (e-mail: somdatta\_goswami@brown.edu).}
\thanks{Xiaoying Zhuang is with the Department of Geotechnical Engineering,College of Civil Engineering, Tongji University, Shanghai, China and is also the Chair of Computational Science and Simulation Technology, Institute of Photonics, Faculty of Mathematics and Physics,Leibniz University, Hannover, Germany. (e-mail: zhuang@iop.uni-hannover.de).}
\thanks{Timon Rabczuk is with the Division of Computational Mechanics and the Faculty of Civil Engineering, Ton Duc Thang University, Ho Chi Minh City, Vietnam. (e-mail: timon.rabczuk@uni-weimar.de).}
}


\maketitle

\begin{abstract}
Modeling fracture is computationally expensive  even in computational simulations of two-dimensional problems. Hence,  scaling up the available approaches to be directly applied to large components or systems crucial for real applications become challenging. In this work. we propose domain decomposition framework for the variational physics-informed neural networks to accurately approximate the crack path defined using the phase field approach.  We show that coupling domain decomposition and adaptive refinement schemes permits to focus the numerical effort where it is most needed: around the zones where crack propagates. No a priori knowledge of the damage pattern is required. The ability to use numerous deep or shallow neural networks in the smaller subdomains gives the proposed method the ability to be parallelized. Additionally,  the framework is integrated with adaptive non-linear activation functions which enhance the learning ability of the networks, and results in faster convergence. The efficiency of the proposed approach is demonstrated numerically with three examples relevant to engineering fracture mechanics. Upon the acceptance of the manuscript, all the codes associated with the manuscript will be made available on Github.
\end{abstract}

\begin{IEEEImpStatement}
	Machine learning techniques have been increasingly used for modeling of engineering problems. In particular, physics informed neural networks (PINNs) have been shown to be a promising approach for discretizing and solving partial differential equations. However, PINNs are best suited for smooth function approximations and have some difficulties dealing with discontinuities and rapidly changing gradients in the solution. Here, we propose a framework for the simulation of nucleation and propagation of cracks under brittle fracture using a subdomain-based phase-field approach. By subdividing the domain into smaller regions and considering an energy minimization formulation, the discontinuous displacements and singular stress fields can be more accurately represented compared to the residual-based formulation.
\end{IEEEImpStatement}


\section{Introduction}

\IEEEPARstart{I}{n} recent years, scientific machine learning has emerged as an essential tool to address domain specific challenges in computational mechanics and extract insights from scientific datasets and governing partial differential equations (PDEs) through innovative methodological solutions. The theory of employing a neural networks to solve an initial and a boundary value problem defined by a PDE was proposed as early as $1990$'s \cite{lee1990neural,lagaris1998artificial}. However, this discovery remained theoretical due to the requirement of enormous compute power. The latest resurrection of neural networks — the deep-learning revolution — is the direct outcome of the modern GPUs that has turned the one-layer networks of the $1960$s and the two- to three-layer networks of the $1980$s into the $10$-, $15$-, and even $50$-layer networks of today. The advent of physics-informed neural networks (PINNs) \cite{raissi2019physics} served as the catalyst for this revolution for computational mechanics engineers, where the hyperparameters of the DNN are optimized by minimizing the residual of the governing PDE, ensuring that the outputs of the network will necessarily satisfy the physics of the problem. Almost during the same time, a variant of PINNs, the variational energy based PINNs (VE-PINNs) \cite{samaniego2020energy}, was proposed to solve the governing PDE in its weak form. Authors in~\cite{chakraborty2021multigoal,chakraborty2022domain} demonstrated a domain adaptation transfer learning approach and goal functional error estimation technique  to solve PDES based on complex geometry. 
In \cite{samaniego2020energy}, the authors showed that for approximating the growth of fracture and to obtain the crack path, it is obligatory to use the variational formulation since fracture is an energy-driven phenomenon. VE-PINNs was able to alleviate the curse of dimensionality with some trade-offs related to the degree of accuracy and relative ease of modeling.

Fracture modeling is a computationally expensive phenomenon as it demands a very fine mesh to resolve the damage region. To design a computationally efficient approach, the same group also proposed an adaptive refinement scheme within the framework of VE-PINNs to locally refine the domain along the path of the growth of the crack \cite{goswami2020adaptive}. To that end, another promising alternative can be the integration of domain decomposition methods with the VE-PINNs framework. The basic idea behind domain decomposition technique is to divide the global domain into subdomains that can be solved independently and then reconnected by interface conditions. These approaches are naturally applicable to the solution of large-scale problems; their goal is to significantly increase the computational efficiency of simulations. Along the same lines and motivated by the extended physics informed neural networks (XPINNs) framework \cite{jagtap2020extended}, in this work we propose a domain decomposition framework for variational energy based XPINNs (VE-XPINNs) to resolve crack paths in brittle materials.

The developed framework employs phase field modeling approach, a popular continuous fracture modeling technique. VE-XPINNs are trained on the governing coupled PDE (variational form) of the phase field approach to encode the vector valued elastic field and the scalar valued phase field based on the initial crack location, material properties and the characteristic width of the crack. One significant advantage of the proposed variational energy formulation is that it requires derivatives one order lower than the conventional residual minimization approach, which results in better computational efficiency. Additionally, motivated by the previous works of V-PINNs \cite{kharazmi2021hp}, we use Gauss quadrature points to evaluate the integrals, over a domain. To begin with, the computational domain is divided into a number of elements and then, the quadrature points are generated within each element, for an efficient integration of non-smooth functions like fracture.

In the following, we summarize the new contributions of the current work:
\begin{itemize}
    \item a generalized domain decomposition framework for variational energy based PINNs to estimate the crack path defined using the phase field approach.
    \item the proposed method employs separate neural networks for each subdomain, which allows the use of tailored NNs and optimizers specifically catering to the needs of a domain, improving computing efficiency and avoiding problems related to overfitting.
    \item the use of adaptive activation functions in this framework allows for better learning and faster convergence of the solution.
\end{itemize}

The remainder of the paper is organized as follows. In \autoref{subsec:phase_field}, we discuss the problem statement for phase-field modeling of brittle fracture using VE-XPINN. Implementation of the domain decomposition method for solving fracture mechanics problems using phase field method is discussed in \autoref{subsec:dd}. Numerical examples illustrating the performance of the proposed approach are presented in \autoref{sec:numericals}. Finally, \autoref{sec:conclusion} presents the concluding remarks.

\section{Methodology}
\label{sec:methodology}

\subsection{Phase-field based fracture modelling}
\label{subsec:phase_field} 

In this section, we briefly put forth the phase field formulation for brittle fracture analysis. The integration of two fields, the vector-valued elastic field and the scalar-valued phase field, is required to model fracture using the phase field approach. While crack nucleation may be influenced by stress, crack propagation necessitates an increase in the fracture energy or surface energy of a solid, $\Psi_c$ \cite{A.Griffith1921}. As a result, the energy criteria is used in the phase field approach to analyse the growth of fracture.

Let us begin with the linear elastic problem on an arbitrary body, $\Omega$, with external boundary $\partial \Omega$. The displacement at each material  point $\boldsymbol{x}$ is denoted by $\boldsymbol u(\boldsymbol{x})$. In addition, we assume a small strain tensor, $\boldsymbol \epsilon$ at each material point defined as,
\begin{equation}
    \boldsymbol \epsilon(\boldsymbol u) =\frac{1}{2}\left(\nabla \boldsymbol u + \nabla \boldsymbol u^T  \right).
\end{equation}
The displacement field satisfies the Dirichlet and Neumann boundary conditions. Considering the material is isotropic and is linearly elastic, the stored elastic energy, $\Psi_c$ at any material point in the body is described by the energy density function, $\Psi_0(\boldsymbol \epsilon)$ as:
\begin{equation}
    \Psi_0(\boldsymbol \epsilon) = \frac{1}{2} \lambda \text{tr}(\boldsymbol{\epsilon})^2+\mu \text{tr}(\boldsymbol{\epsilon}^2),
\end{equation}
where \text{tr}$(\cdot)$ denotes the trace of the strain tensor and $\lambda$ and $\mu$ are the Lam\'e constants. The Cauchy stress tensor, $\boldsymbol \sigma$, at any material point on the domain $\Omega$ can be computed as:
\begin{equation}
  \boldsymbol \sigma = \partial_{\boldsymbol{\epsilon}}\Psi_0(\boldsymbol\epsilon) = \lambda \text{tr}(\boldsymbol\epsilon) \boldsymbol{I} + 2 \mu \boldsymbol \epsilon. 
\end{equation}
The momentum-balance equation for the elastic field, considering an isotropic solid, can be written as:
\begin{equation}\label{eq0}
    \begin{split}
    &\nabla \cdot \boldsymbol\sigma  = \boldsymbol f(\boldsymbol{x}) \quad \quad \,\;\;\text{in } \Omega,\\
    &\boldsymbol u =\overline{\boldsymbol u} \qquad \qquad \;\;\;\;\;\; \text{on } \partial \Omega_D,\\
    &\boldsymbol\sigma \cdot n = \boldsymbol{t}_N\qquad \,\;\;\;\;\;\;\text{on } \partial \Omega_N,
  \end{split}
\end{equation}
where the Dirichlet and Neumann boundaries are represented by $\partial\Omega_{D}$ and $\partial\Omega_{N}$, respectively, $\boldsymbol{t}_N$ is the prescribed boundary force and $\boldsymbol{\overline u}$ is the prescribed displacement. The stored internal potential energy of the body, for homogeneous Neumann boundary conditions, is given by:
\begin{equation}
    \Psi_{int}(\boldsymbol u) =\int_{\Omega} \Psi_0(\boldsymbol\epsilon) \,d \boldsymbol{x} ,
\end{equation}
and the external energy is given by,
\begin{equation}
    \Psi_{ext}(u)= \int_{\Omega} \boldsymbol f\cdot \boldsymbol u ~\,d \boldsymbol{x} + \int_{\partial \Omega_N} \boldsymbol{t}_N \cdot \boldsymbol u ~\,d\gamma,
\end{equation}
where $\boldsymbol{f}$ is the body force. Therefore, the optimization problem for linear elasticity is defined as:
\begin{equation}
    \min: \Psi_e = \Psi_{int}-\Psi_{ext},
    \label{eq3}
\end{equation}
subjected to the pre-defined boundary conditions. 

The equilibrium equation for the elastic field for an isotropic model, considering the evolution of crack, involves the degradation of the stiffness of the material around the area of the crack by penalizing the Cauchy stress tensor, $\boldsymbol \sigma$ with a monotonically decreasing stress-degradation function, $g(\phi)$. A common form of the stress-degradation function in the literature is \cite{miehe2010thermodynamically} $$ g(\phi) =(1-\phi)^2.$$ With evolving damage, only the tensile component of the principal stress degrades while the compressive component remains invariant \cite{miehe2010phase}. Therefore, in order to prevent the growth of crack inside regions under compression, a tension-compression split of $\Psi_0(\boldsymbol{\epsilon})$ is considered. With evolving damage, $g(\phi)$ is applied only to the tensile component of the principal strain. Consequently, due to the growth of fracture, the anisotropic constitutive assumption for the degradation of the elastic strain energy, can be stated as:
\begin{equation}
   \Psi_e(\boldsymbol{\epsilon}):= g(\phi) \Psi_0^{+}(\boldsymbol{\epsilon}) + \Psi_0^{-}(\boldsymbol{\epsilon}),
\end{equation}
where $\Psi_0^{+}(\boldsymbol{\epsilon})$ and $\Psi_0^{-}(\boldsymbol{\epsilon})$ denote the strain energies computed from the positive and negative components of the principal strains, respectively. 

\noindent The governing equation for the phase-field is written as \cite{amor2009regularized,bourdin2011time}:
\begin{equation}\label{eq:phasefield_eq}
       \frac{G_c}{l_0}\phi - G_{c}l_{0}\nabla^{2}\phi = -g'(\phi)\mathcal H(\boldsymbol{x},t) \text{ on } \Omega,
\end{equation}
where $G_c$  represents the critical energy release rate (property of material), $\mathcal H(\boldsymbol{x},t)$ is a local strain-history function, $\ell_0$ is the length scale parameter that controls the width of the diffusive region of the crack. The effect of $l_0$ has been verified by a series of numerical simulations \cite{zhou2018phase,zhou2018pha}, demonstrating that the crack region has a larger width with an increasing $\ell_0$ while the phase field represents a sharp crack topology when $\ell_0 \to 0$. The phase field is used to smear out the crack surface over the domain $\Omega$. In the regularized model, correspondingly, the phase field must satisfy the following condition:
\begin{equation}
    \phi(\boldsymbol{x},t) =
\left\{
	\begin{array}{ll}
		0  & \mbox{the material is intact, }  \\
		1 & \mbox{the material is completely cracked. } 
	\end{array}
\right.
\end{equation}
The local strain-history functional, $\mathcal{H}(\boldsymbol{x},t)$ contains the maximum positive tensile energy, $\Psi_0^{+}$ in the history of deformation of the system \cite{miehe2010thermodynamically} and is defined as:
\begin{equation}
    \mathcal{H}(\boldsymbol{x},t)=\max_{s\in [0,t]}\Psi_0^{+}(\boldsymbol \epsilon(\boldsymbol{x},s)).
\end{equation}
The strain-history function can also be used to initialize or nucleate the crack \cite{miehe2010phase}. In particular, we set the initial strain-history function as \cite{Borden2012}:
\begin{equation}
 \mathcal{H}(\boldsymbol{x},0) =
\left\{
	\begin{array}{ll}
		\frac{BG_c}{2 \ell_0}\left(1- \frac{2 d(\boldsymbol x)}{\ell_0}\right)  & \mbox{if } d(\boldsymbol x)\le \frac{\ell_0}{2}, \\
		0 & \mbox{otherwise. } 
	\end{array}
\right.
\label{eq5}
\end{equation}
where $d(\boldsymbol x)$ is the distance from $\boldsymbol x$ to the crack tip, $B$ is a scalar parameter controlling the magnitude of the scalar history field and is computed as:
\begin{equation}
    B(\phi) =\frac{1}{1-\phi}\;\; \mbox{for } \phi<1.
\end{equation}
The fracture energy, $\Psi_c$ of the newly formed cracks is expressed as:
\begin{equation}
    \Psi_c =\int_{\Omega}\left(G_c\Gamma_n(\phi)+g(\phi) \mathcal{H}(\boldsymbol{x},t)\right) \,d \Omega.
\end{equation}
Here, $\Gamma_n(\phi)$ represents the crack density functional and $n$ is the order of the corresponding phase field model. For the second-order phase field model, $n = 2$, while for the fourth-order phase field model, $n=4$. The fourth order phase model includes higher-order derivatives of $\phi$, leading to greater regularity in the exact solution of the phase field. In this case the cracked surface can be captured more accurately and fewer degrees of freedom are required to resolve the crack path relative to the second-order model. For the second and fourth order phase field model the crack density functionals, $\Gamma$ are defined as \cite{borden2014higher}:
\begin{equation}
    \begin{split}
   \Gamma_2(\phi) & = \frac{1}{2\ell_0} \int_{\Omega} \left(\phi^2+\frac{\ell_0^2}{2} |\nabla \phi|^2\right) \,d \Omega, \\
   \Gamma_4(\phi) & = \frac{1}{2\ell_0} \int_{\Omega} \left(\phi^2+\frac{\ell_0^2}{2} |\nabla \phi|^2 + \frac{\ell_0^4}{16} |\Delta \phi|^2\right) \,d \Omega.
   \end{split}
\end{equation}
In this work we study the growth of fracture employing the variational energy based XPINN approach. To that end, we solve the optimization problem defined as \cite{A.Griffith1921, Bourdin2000, Borden2014}:
\begin{equation}\label{eq4}
    \begin{split}
  \textrm{Minimize }&: \mathcal E= \Psi_e+\Psi_c, \\
  \textrm{constrained to }&: \boldsymbol u = \bar{\boldsymbol u}~~\text{on } \partial \Omega_D, \\
  \textrm{such that }& : \Psi_e = \int_{\Omega} \left( g(\phi) \Psi_0^{+}+\Psi_0^{-}\right) \,d \Omega, \\
  & \;\;\;\;  \Psi_c = \int_{\Omega} G_c \Gamma_n+ g(\phi) \mathcal{H}(\boldsymbol{x},t) \,d \Omega.
   \end{split}
\end{equation}
The homogeneous Neumann boundary conditions are automatically satisfied when the variational energy principle is used. 

\subsection{Variational energy based XPINNs}
\label{subsec:dd}

When performing a traditional numerical model analysis, like FEM and IGA, to simulate the mechanical response of solids under loading dynamics and crack propagation, as described in \autoref{subsec:phase_field}, the resulting computational burden can be very high, especially when addressing a fully three-dimensional problem. In fact, a very refined mesh is required to properly describe the discrete crack boundaries and resolve the crack tip. This results in a large number of degrees of freedom that needs to be solved to compute the quantities of interest. Hence, the high computational burden of a three-dimensional simulation remains a major disadvantage. To alleviate the burden, in this section we introduce the domain decomposition method.

Suppose the computational domain $\Omega$ is partitioned into $p$ subdomains $\{\Omega_s:s=1,2\ldots,p\}$ such that $\Omega= \bigcup_{s=1}^p \Omega_s$. By partitioning, we imply that the division is non overlapping, and the subdomains intersect only on their common boundaries which is also called as \emph{interface}. To simplify the discussion, consider the linear elastic problem defined in \autoref{eq0} as an example, where on a particular subdomain $\Omega_s$ we have:
 \begin{equation}\label{eq1}
-\nabla \cdot \boldsymbol\sigma_s(\boldsymbol{x})  = \boldsymbol f_s(\boldsymbol{x})\;\; \text{in } \Omega_s, 
\end{equation}
where $\boldsymbol f_s$ is the body force in the $s$-th subdomain, and $\boldsymbol \sigma_s$ is the Cauchy stress tensor associated with the material point in the $s$-th domain. Each subdomain is constrained by the continuity condition:
\begin{equation}
\boldsymbol u_{i}(\boldsymbol{x})= \boldsymbol u_{j}(\boldsymbol{x})\qquad \text{on } \Gamma_{ij},  \\
\end{equation}
where  $\Gamma_{ij}$ is the interface between the $i$-th and $j$-th subdomains, and $\boldsymbol u_i$ and $\boldsymbol u_j$ are the displacements on $\Omega_i$ and $\Omega_j$ respectively. The above boundary conditions also represent the interface transmission conditions between the neighbouring subdomains. For each of these subdomains, a separate deep neural network (DNN) is deployed.

A DNN \cite{chakraborty2021multigoal}, $\mathfrak{N}_{\mathrm{L}}:\mathbb{R}^d \mapsto \mathbb{R}^p$ is a non linear function defined as  concatenations of affine maps with point-wise non linearities, in the form:
\begin{equation}
     \mathfrak{N}_{\mathrm{L}}(\boldsymbol{x})=\mathcal{W}_{\mathrm{L}}  \tau_{\mathrm{L}} \left(\ldots  \tau_3 \left( \mathcal{W}_{2} \tau_2 \left(\mathcal{W}_1 \boldsymbol{x} +\bold{\beta}_1 \right) +\bold{\beta}_2 \right) \ldots \right) +\bold{\beta}_{\mathrm{L}}  
\end{equation}
where, $\mathcal{W}_i$'s are weight matrices may not necessarily be square and $\beta_i$'s are referred as bias vectors. $\mathrm{L}$ is the depth (the number of layers) of the network and $\mathrm{L}\geq 3$. The input vector is $\boldsymbol{x}$ and the output vector of any arbitrary $\ell$-th layer is denoted by $\mathfrak{N}_{\ell}(\boldsymbol{x})$ in particular,  $\mathfrak{N}_{0}(\boldsymbol{x})=\boldsymbol{x}$. The non-linear monotonic function $\tau$ is known as activation function applied layer-wise to any vector, however the dimension of the input vector may vary depending on layers. However, regardless of input dimension, the activation function performs the same operations on all input entries. The activation function in the final layer is linear.

Recently, the implementation of adaptive activation functions \cite{kunc2021transformative,jagtap2020adaptive} (AAF) have been receiving more attention, i.e, function possess  trainable parameters that changes their shape in the training process. AAFs have shown advantages such as reducing training times while significantly improving the rate of convergence and increased accuracy of the solution. Therefore the model ends up with better learning capabilities than the the model with a fixed activation function. To expand upon these notions, an adaptive model that can easily be linked with existing methodologies to increase overall accuracy of VE-XPINNs has been investigated in this work. The simplest form is just to add a parameter to a particular neural network that controls one of its properties, e.g. slope. Following the approach in \cite{jagtap2020adaptive}. in this work we introduced two parameters $\mathfrak{c}$ (scaling parameter) and $\alpha$ (hyperparameter to be optimized) in the activation function as:
\begin{equation}
    \tau_{\mathrm{L}}(\ldots\mathfrak{c} \cdot \alpha~\tau_3(\mathfrak{c} \cdot \alpha~\cdot\tau_2(\cdot))).
\end{equation}
The primary goal of fracture mechanics is to determine the crack path. In this work, we have studied displacement-controlled fracture, \textit{i.e.} this is growth of the crack by applying a displacement increment until failure. In addition, we assumed a constant displacement step, $\Delta u$. The proposed VE-XPINN is trained and the strain-history function of each sub-domain is updated at each displacement increment in this configuration. In the beginning, the network parameters are initialized using a Gaussian distribution with Xavier initialization technique \cite{glorot2010understanding}. Let us first use a DNN to represent the displacement field, $\boldsymbol U_s$, and the phase-field, $\phi_s$ of the $s$-th subdomain, such that
\begin{equation}\label{eq:primal_var}
    \left( \boldsymbol U_s, \phi_s \right) = \mathfrak{N}_s\left(\boldsymbol x_s; [\mathcal{W}_s, \beta_s]\right).
\end{equation}
Each subdomain is discretised into $n_e^s$ elements, and we generate the Gauss points within each element and their corresponding weights. The total variational energy is calculated at the Gauss points to approximate the integral. The displacements and the phase field at the common interfaces are assumed to be continuous. Therefore, for solving the PDE as a whole, we defining the loss function separately for the interior of each subdomain and the interfaces as follows:
\begin{equation}\label{eq2}
\begin{split}
  \mathcal{L}_s^{int}(\mathcal{W}, \beta_s)&=  \Psi_e(\epsilon(U_s), \phi_s) + \Psi_c(G_c, \boldsymbol x_s) \\
  \mathcal{L}_{ij}^{iface}(\mathcal{W}, \beta_s) &= 
  \frac{W_1}{N_{\Gamma_{ij}}} \sum_{k=1}^{N_{\Gamma_{ij}}}|\boldsymbol u_i(\boldsymbol{x}_k) - \boldsymbol u_j(\boldsymbol{x}_k)|^2 \nonumber \\
  & +\frac{W_2}{N_{\Gamma_{ij}}} \sum_{k=1}^{N_{\Gamma_{ij}}}\left |\phi_i(\boldsymbol{x}_k)- \phi_j(\boldsymbol{x}_k)\right|^2 \nonumber
\end{split}  
\end{equation}
where $N_{\Gamma_s}$ represents the total training points scattered over its interfaces, respectively. The interface condition endowed with the loss function plays an important role for stitching the subdomains together. It ensures that the data information should propagate among the neighbouring domains. A sufficient number of collocation points needs to be generated on the interface leading to faster convergence of the algorithm. In \autoref{eq2}, $W_i$ $\forall\;\;i = {1, 2}$ are penalizing parameters, which are chosen manually for balancing each of the loss terms and also for faster convergence. From the viewpoint of optimization  it could be more sophisticated to multiply the constraints with suitable penalty coefficients. A method for adaptively choosing these penalty parameters has been proposed in \cite{mcclenny2020self}. 

The analysis of fracture requires a very fine mesh in the vicinity of the crack, to capture local quantities of interest. To that end, we could either pre-refine the expected path of crack growth based on prior information available from the literature, or need an efficient adaptive refinement scheme that generates more quadrature points along the path that the crack grows. Since knowledge to pre-refine the domain is strictly limited, in this work we propose an adaptive $\it{h}$-refinement scheme for VE-XPINN to enhance the formulation. The refinement scheme locally refines the crack path as the crack grows. 

One of the most difficult challenges for adaptive refinement schemes is determining the appropriate refinement criterion, which determines the location and level of refinement. The resulting refinement scheme has two components: (i) a model training status derived from the loss function of each subdomain that serves as an error indicator, and (ii) an algorithmic component that performs domain refinement. Since the values of the phase-field variable $(\phi)$ increase from zero to one from the intact domains to the crack surface, both the phase field variable and its gradient are used as error estimators to determine the intensity of training points around crack surfaces. Initially the domain is divided into several subdomain and we consider a threshold $\phi_{thres}$ and hence the expression $\phi>\phi_{thres}$ is used as a refinement criteria to improve the domain  refinement around crack surfaces especially the areas near the crack-tips. The region chosen for refinement should not be finer than a chosen maximum refinement level. On the other hand, for the phase field area  a recovery-based posteriori error estimator~\cite{zienkiewicz1992superconvergent} is used to drive adaptivity. The refinement strategy is based on refining the region that contributes the most to error. The process of generating fine refinements in order to achieve optimal results in which the error norm is distributed evenly across the decomposed subdomains. Precisely, the underlying idea is to minimize the loss function in each subdomain, prior to the resulting error function meets a predefined tolerance. 
 
\section{Numerical Examples}
\label{sec:numericals}

As a demonstration of the effectiveness of our proposed model, we study several benchmark problems from fracture mechanics in this section. The first example illustrates a one-dimensional elastic bar with a crack in the center subjected to sinusoidal loading. An analytical solution to this problem exists. Thus, the proposed adaptive refinement approach can be validated. In addition, we have considered two other examples; the  single-edge notched tension test and a specimen with an eccentric hole under tension. All the experiments have been carried out using the \emph{TensorFlow} framework on Google Colab Pro GPU.

\subsection{One dimensional cracked elastic bar}
\label{subsec:oneD}
\begin{figure}[htbp!]
    \centering
    \includegraphics[width = 0.45\textwidth]{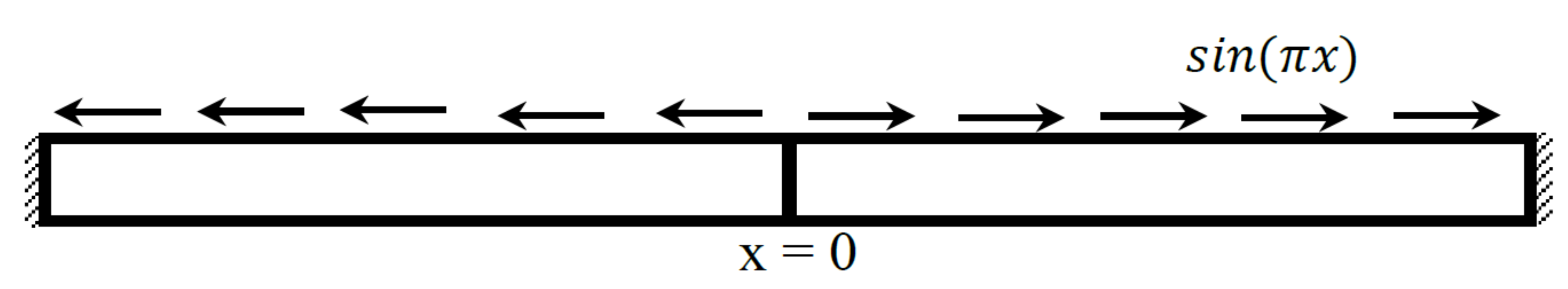}
    \caption{Geometrical setup of one-dimensional elastic bar with crack.}
  \label{fig:1D_setup}
\end{figure}

We consider a one-dimensional bar that is fixed at both the ends $(x=-1,x=1)$ and is subjected to a sinusoidal load. The bar has a crack at the center, $x=0$. The geometrical setup is presented in \autoref{fig:1D_setup}. 
For simplicity $E$ is assumed to be unity and the strain ($\epsilon$) is assumed to be non-negative in the crack zone. Hence, the stress-strain relation is obtained by: 
\begin{equation}
    \sigma = (1-\phi)^2\epsilon.
\end{equation}
The crack at center is imposed by the following step function:
\begin{equation}
    \mathcal{H}(x,0):= \left\{
	\begin{array}{ll}
		1000  & \mbox{if } d(x) \leq \ell_0 \\ 
		0 & \mbox{if } d(x) > \ell_0,
	\end{array}
\right.
\end{equation}
where $\ell_o$ is known as length-scale parameter. In particular we choose $\ell_0=\frac{1}{80}$. The displacement field satisfies Dirichlet boundary condition, \i.e,
\begin{equation}
    u(-1)=u(1)=0.
\end{equation}
The analytical solution for the displacement field \cite{schillinger2015isogeometric} is discontinuous and is given by,
\begin{equation}\label{eq:exact_u}
    u_{ex}:= \left\{
	\begin{array}{ll}
		\frac{1}{\pi^2} \sin (\pi x)-\frac{1+x}{\pi}  & \mbox{if } x<0 \\ 
		\frac{1}{\pi^2} \sin (\pi x)+\frac{1-x}{\pi} & \mbox{if } x \geq 0
	\end{array},
\right.
\end{equation}
and solution for the phase field~\cite{miehe2010thermodynamically} is,
\begin{equation}\label{eq:exact_phi}
    \phi_{ex}:=\exp\left(\frac{-|x|}{\ell_0}\right).
\end{equation}

In this example, we have considered two possibilities. The first setup uses two subdomains such that
\begin{equation*}
    \underbrace{[-1.0,1.0]}_{\Omega}= \underbrace{[-1.0,0.0]}_{\Omega_1}\cup\underbrace{[0.0,1.0]}_{\Omega_2}, 
\end{equation*}
while the second setup employs the computational domain $\Omega$ is subdivided into four subdomains, such that
\begin{equation*}
    \underbrace{[-1,1]}_{\Omega}= \underbrace{[-1,-0.5]}_{\Omega_1}\cup\underbrace{[-0.5,0]}_{\Omega_2}\cup\underbrace{[0,0.5]}_{\Omega_3}\cup\underbrace{[0.5,1]}_{\Omega_4}.
\end{equation*}
We have used adaptive \textit{tanh} activation in the hidden layers whereas for the final layer, linear activation function has been considered and $\epsilon_{tol} = 1e-12$. The learning rate $\alpha = 0.001$ is used in the \emph{BFGS} optimizer following \emph{Adam}. The region to be refined is decided based on the critical threshold, $\phi_{thres}$ and percentage of total estimated error for elasticity, $\rho$. In this problem, $\rho = 25\%$ and $\phi_{thres}$ is fixed at 0.2. To ensure that the neural network outputs exactly satisfies the Dirichlet boundary conditions, we set the following:
\begin{equation*}
    \boldsymbol u=[(x+1)(x-1)]\boldsymbol u_{\theta},
\end{equation*}
where $\boldsymbol u_{\theta}$ is the displacement field obtained as output from the neural network. In the proposed VE-XPINN approach, the network is trained by minimizing the total variational energy of the system as defined in \autoref{eq4}. The adaptive refinement scheme generates more quadrature points in the vicinity of the crack. For the setup with four subdomains, the refinement is carried out in the intervals $\Omega_2$ and $\Omega_3$. We start training the network with $100-250-250-100$ Gauss quadrature points and to achieve a relative $L_2$ prediction error of less than $1.5\%$ in $\phi$, the domain is refined adaptively and we finally end the training with $250-600-600-250$ quadrature points in the four domains, respectively. To quantify the accuracy of the results obtained using the proposed approach, the relative $L_2$ error corresponding to $u$ and $\phi$ are reported in \autoref{tab:prob1}. To illustrate the superiority of using the fourth-order phase field model over the second-order model, we simulate the second order model and report a prediction error of $3.4\%$ for $u$ and $2.51\%$ for $\phi$.

\begin{figure}[htbp]
\centering
\subfloat[]{\includegraphics[width=1.5in,height=1.2in]{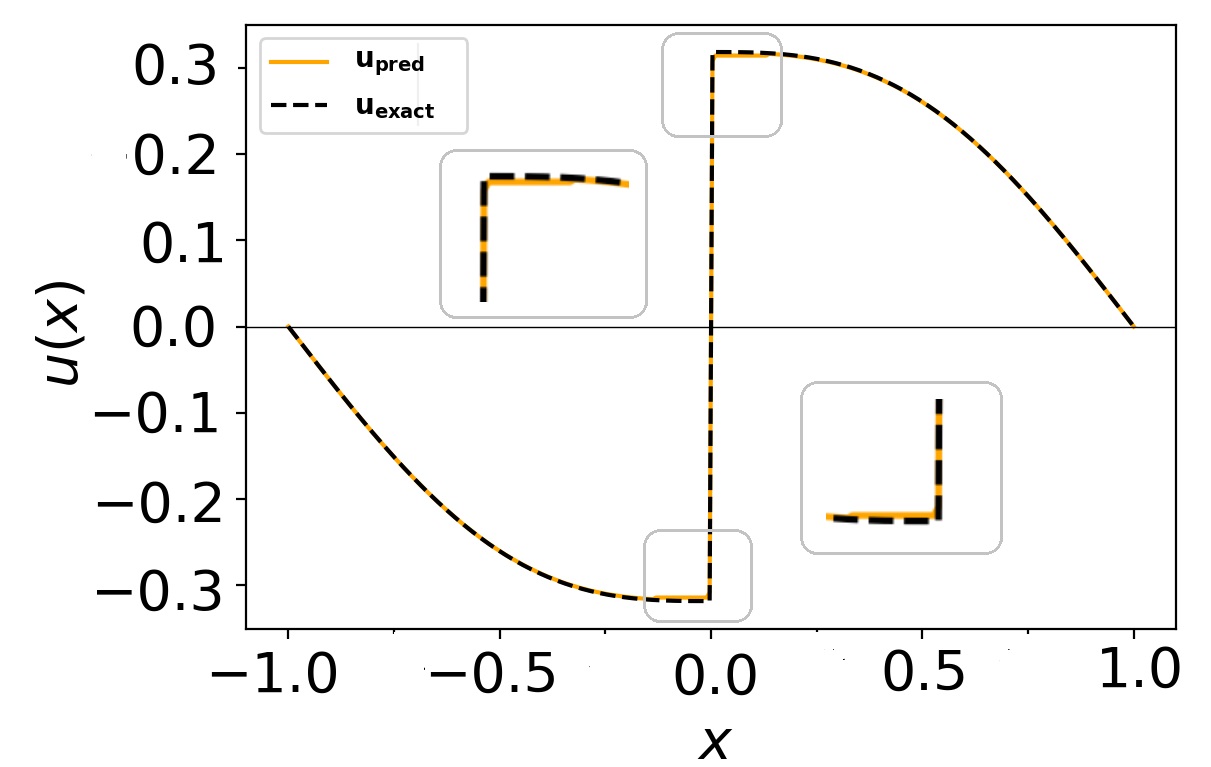}}\quad
\subfloat[]{\includegraphics[width=1.5in,height=1.2in]{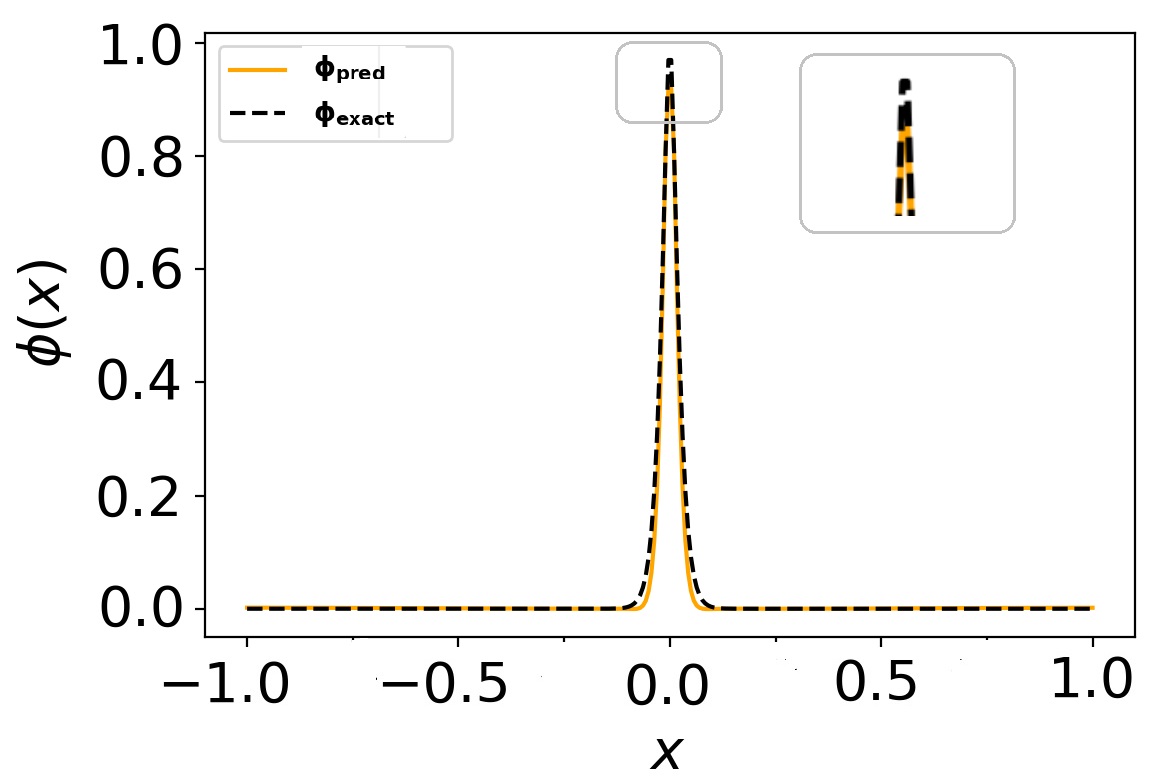}}\quad
\subfloat[]{\includegraphics[width=1.5in,height=1.2in]{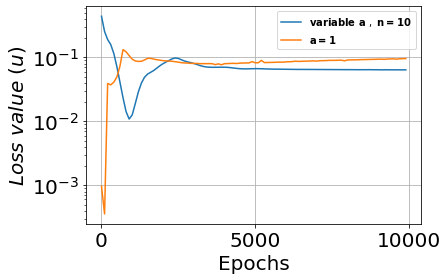}}\quad
\subfloat[]{\includegraphics[width=1.5in,height=1.2in]{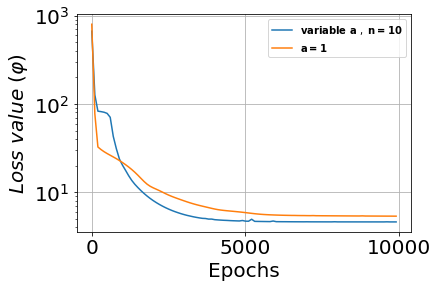}}\quad
\subfloat[]{\includegraphics[width=1.5in,height=1.2in]{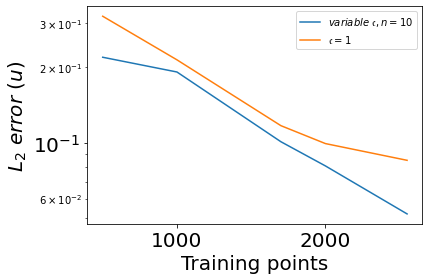}}\quad
\subfloat[]{\includegraphics[width=1.5in,height=1.2in]{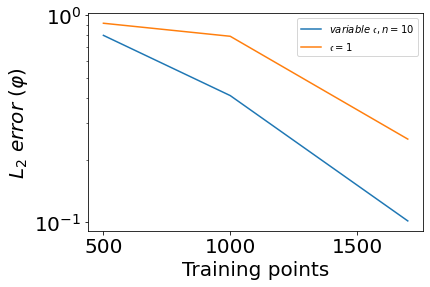}}\\
\subfloat[]{\includegraphics[width=1.5in,height=1.2in]{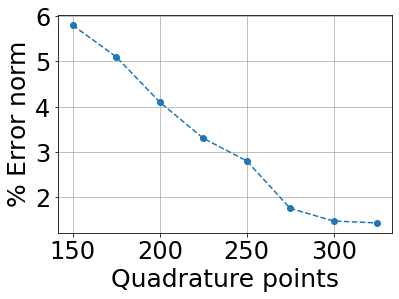}} 
\caption{Results for one-dimensional elastic bar with crack using the domain decomposition framework with the variational energy based PINN (proposed approach). (a) and (b) show the agreement of the proposed approach with the exact solution computed using \autoref{eq:exact_u} and \autoref{eq:exact_phi}. (c)-(f) show the loss plot and the $\mathcal L_2$ error plots for an adaptive activation function vs a constant activation function. Finally, (g) shows the convergence of the solution with the increase in the number of quadrature points.}
\label{figp2}
\end{figure}

\begin{table}[htbp!]
    \caption{Summary of results corresponding to \autoref{subsec:oneD}, when the domain is discretized in $2$ and $4$ subdomains and using the fourth-order phase field model.}
    \label{tab:prob1}
    \begin{tabular}{ccccc}
        \hline
        \textbf{\# of} & \multirow{2}{*}{\textbf{DNN Architecture}} & \textbf{Integration} & \multicolumn{2}{c}{\textbf{Prediction error}} \\ \cline{4-5}
         \textbf{domains} & & \textbf{points} & $\mathcal L_2^{rel,u}$ & $\mathcal L_2^{rel,\phi}$ \\ \hline
         $2$ & $\left[1,10,10,10,2\right]$ & $800,800$ & $3.4\%$ & $2.51\%$ \\
         $4$ & $\left[1,10,10,10,2\right]$ & $250,600,600,250$ & $1.19\%$ & $<1\%$ \\
         \hline
    \end{tabular}
\end{table}
\autoref{figp2} (a) and (b) depicts the solution plots of $\boldsymbol u$ and $\phi$ using the fourth-order phase field model with $4$ subdomains. As expected, in \autoref{figp2} (g), we see a sharp decrease in the percentage error with an increase in the number of quadrature points in the vicinity of the crack. In \autoref{figp2} (c)-(f), we present the loss plots and the relative $\mathcal L_2$ error plots obtained during the training for a framework integrated with the adaptive activation function and also with a constant activation function. The plots show that the simulation with adaptive activation function converges to a lower $\mathcal L_2$ error, which shows that the adaptive activation function provides advantages like an increase in accuracy and fast convergence rate, especially in the early training period.  We noted that for unity scaling factor (when $\mathfrak{c}=1)$ the optimization process is slower but the tuning process speeds up once the scaling factor increases from $\mathfrak{c}=1,\ldots 10$. The inclusion of scalable hyperparameter dynamically changes the topology of loss values and thereby attaining faster convergence towards the minima.

Finally, we compare our results with one of the earlier work from this group on adaptive fourth-order phase field using VE-PINNs \cite{goswami2020adaptive}. The work reports a relative $\mathcal L_2$ error of $2.19\%$ and is $1.6\%$ of $\boldsymbol u$ and $\phi$, respectively. With these results, we summarize that the proposed approach approximates the solution better than the work in the existing literature.

\subsection{Single-edge notched tension example}
\label{sec:2D_tension_Test}

\begin{figure}[htbp!]
    \centering
    \subfloat[]{\includegraphics[width = 0.45\linewidth]{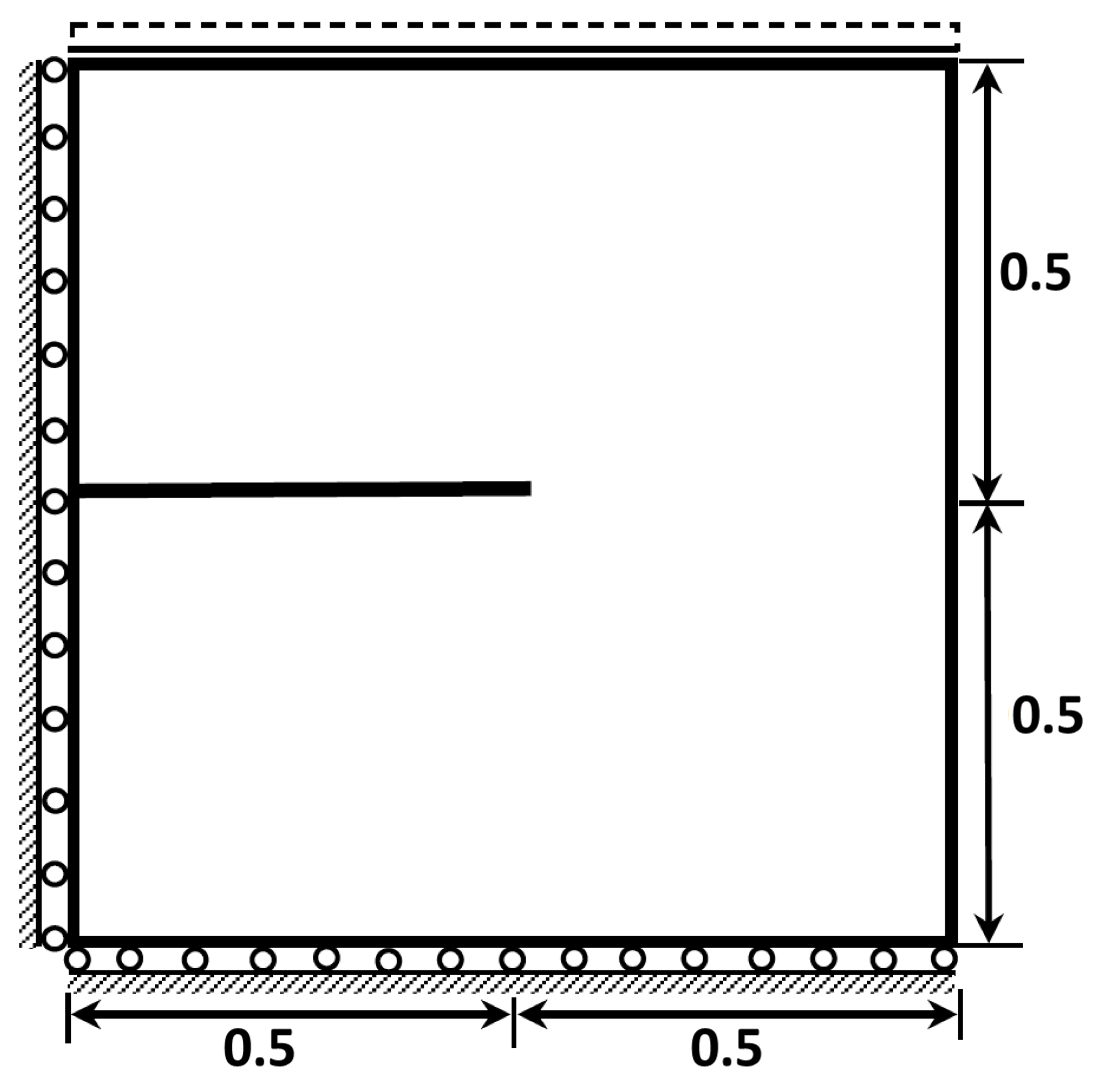}}
    \subfloat[]{\includegraphics[width = 0.45\linewidth, height = 0.45\linewidth]{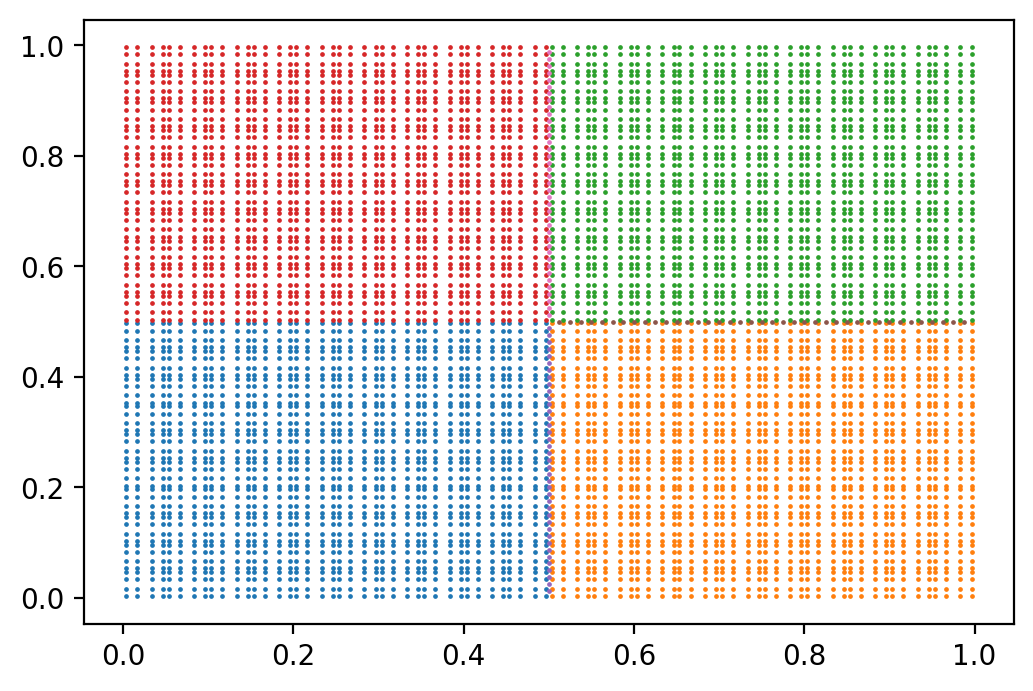}}\quad 
    \subfloat[]{\includegraphics[width = 0.45\linewidth, height = 0.45\linewidth]{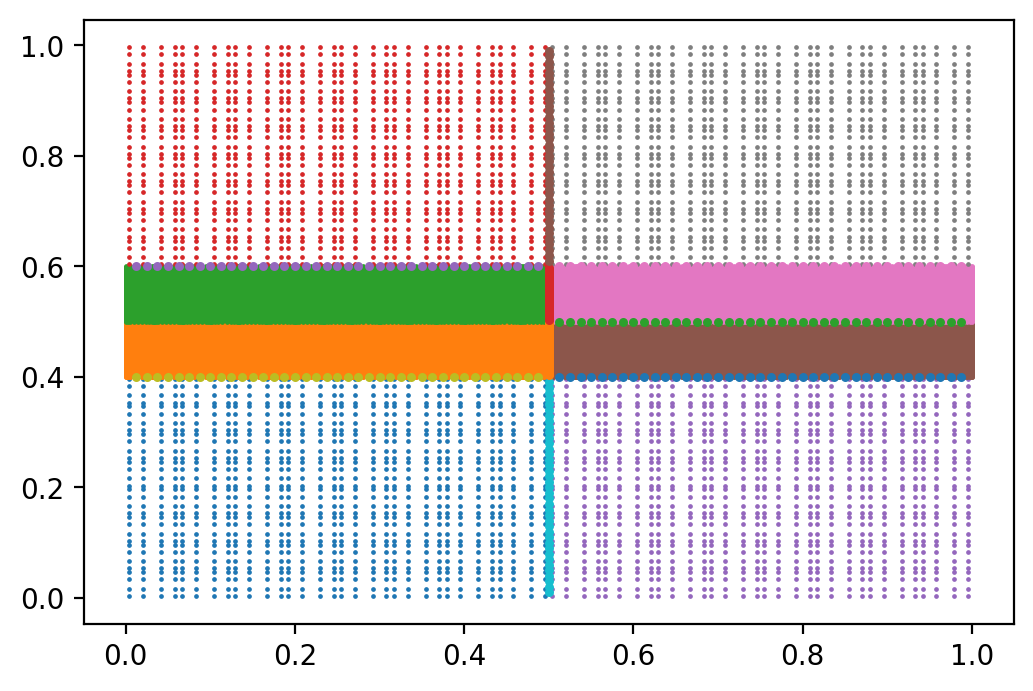}}\quad
    \subfloat[]{\includegraphics[width = 0.45\linewidth, height = 0.45\linewidth]{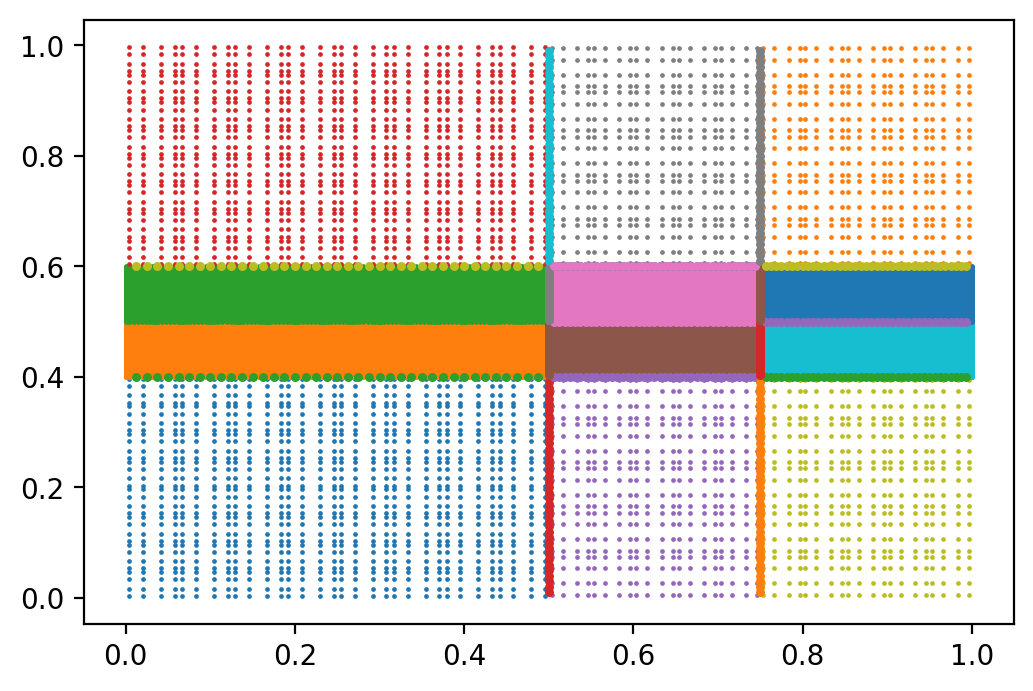}}
    \caption{(a) Geometrical setup and boundary conditions for the single-edge notched tension test. All the units are in mm. (b)-(d) Distribution of training data over four subdomains, eight subdomains and twelve subdomains.}
    \label{fig:setup}
\end{figure}

\begin{figure}[htbp]
\centering
\subfloat[] {\includegraphics[width=1in,height=1in]{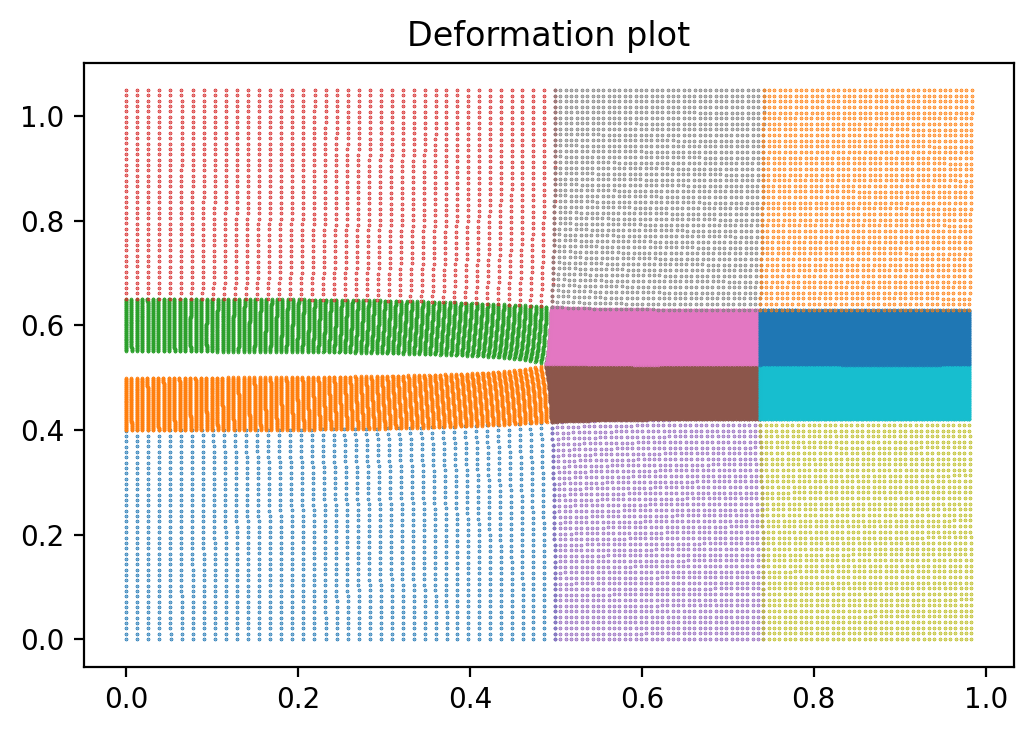}}\quad
\subfloat[] {\includegraphics[width=1in,height=1in]{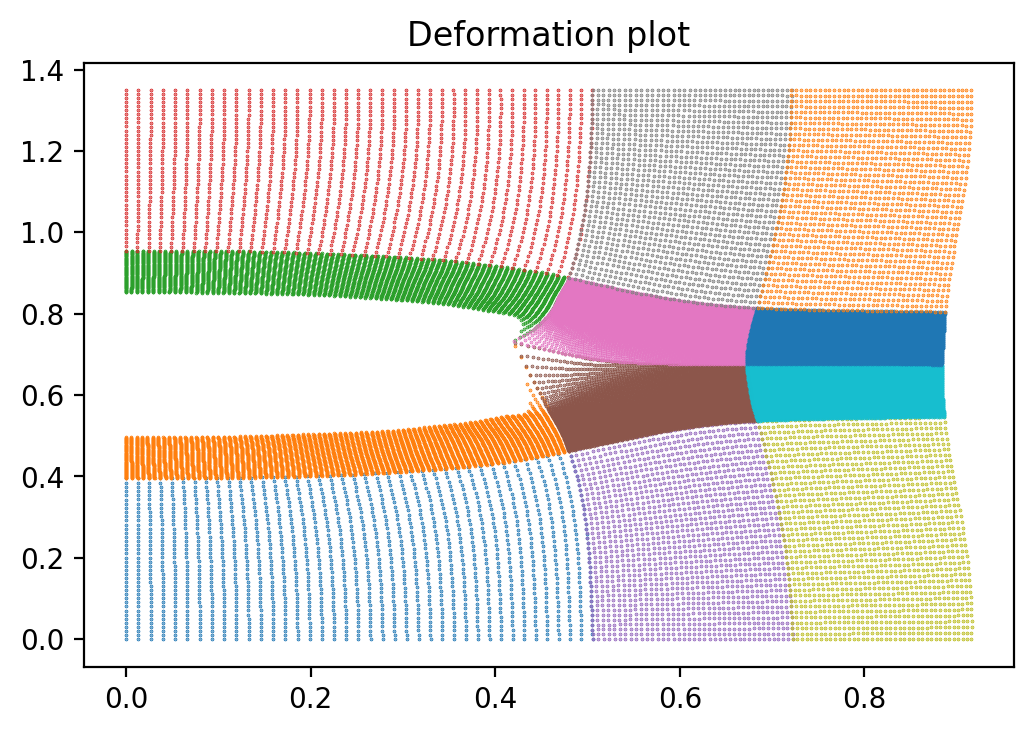}}\quad
\subfloat[] {\includegraphics[width=1in,height=1in]{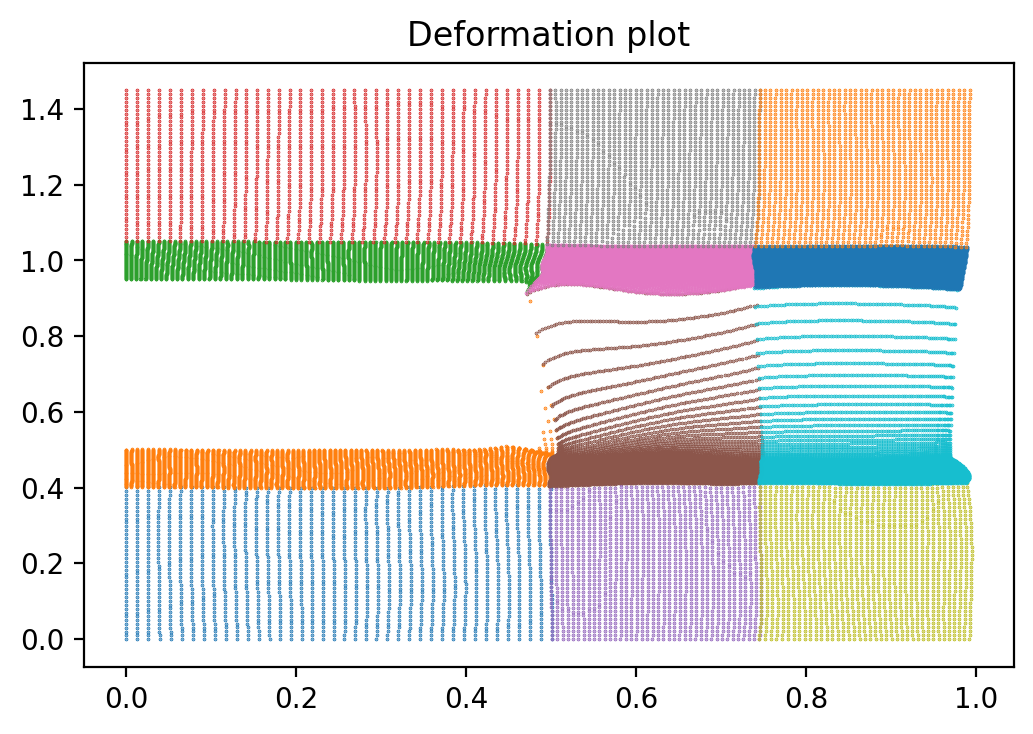}}\quad
\subfloat[] {\includegraphics[width=1in,height=1in]{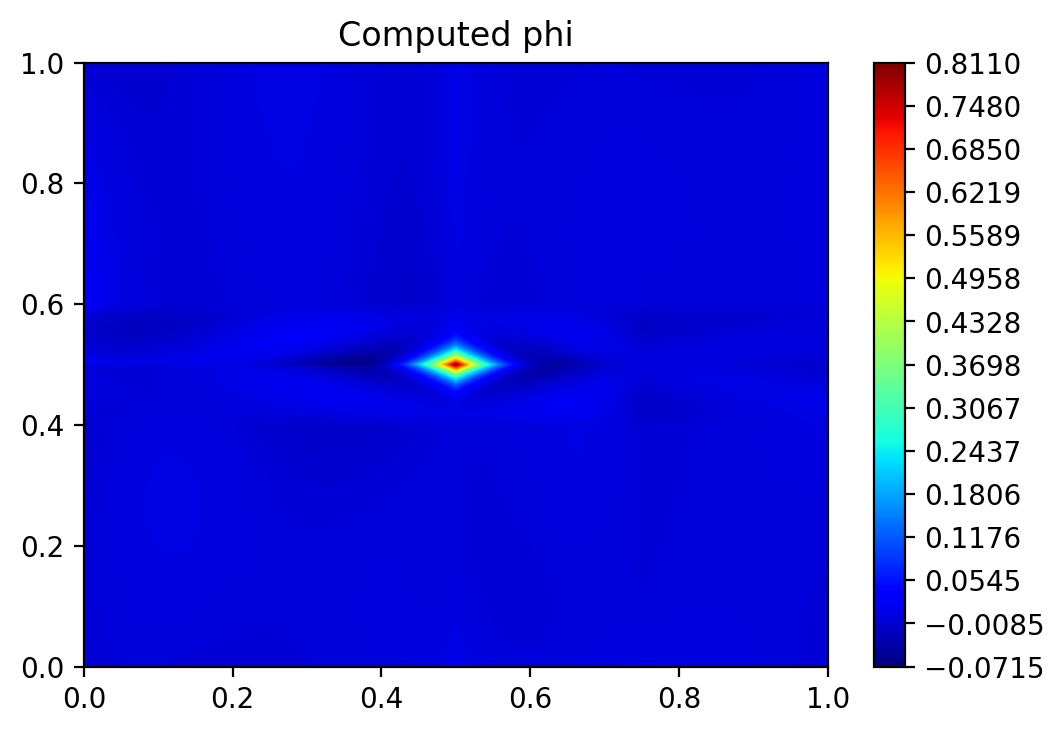}}\quad 
\subfloat[] {\includegraphics[width=1in,height=1in]{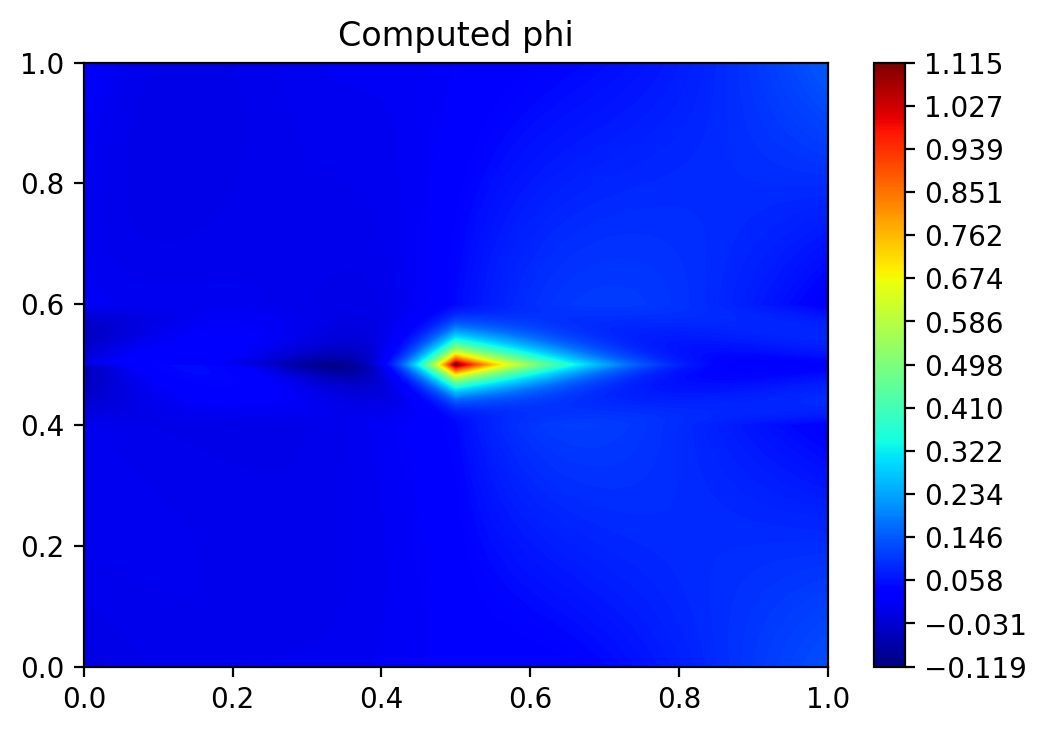}}\quad
\subfloat[] {\includegraphics[width=1in,height=1in]{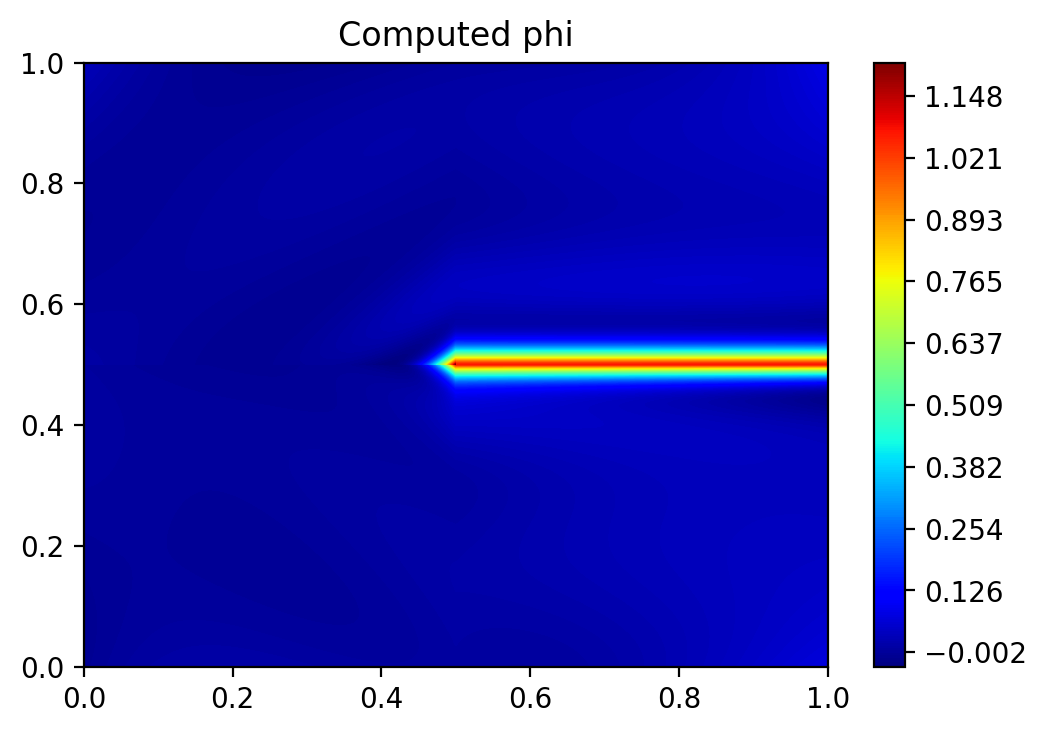}}
\caption{The experiment with $12$ subdomains with adaptive refinement for prescribed displacements of (from (a) to (c)) $10^{-3}$,\;$7\times 10^{-3}$ and $10^{-2}$  mm using the fourth-order phase field model in the domain decomposition framework of variational energy based XPINN approach.}
\label{figp4}
\end{figure}

\begin{figure}[htbp]
\centering
\subfloat[]{\includegraphics[width=1in,height=1in]{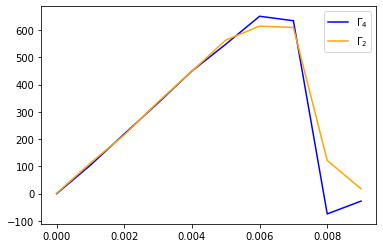}}\quad
\subfloat[]{\includegraphics[width=1in,height=1in]{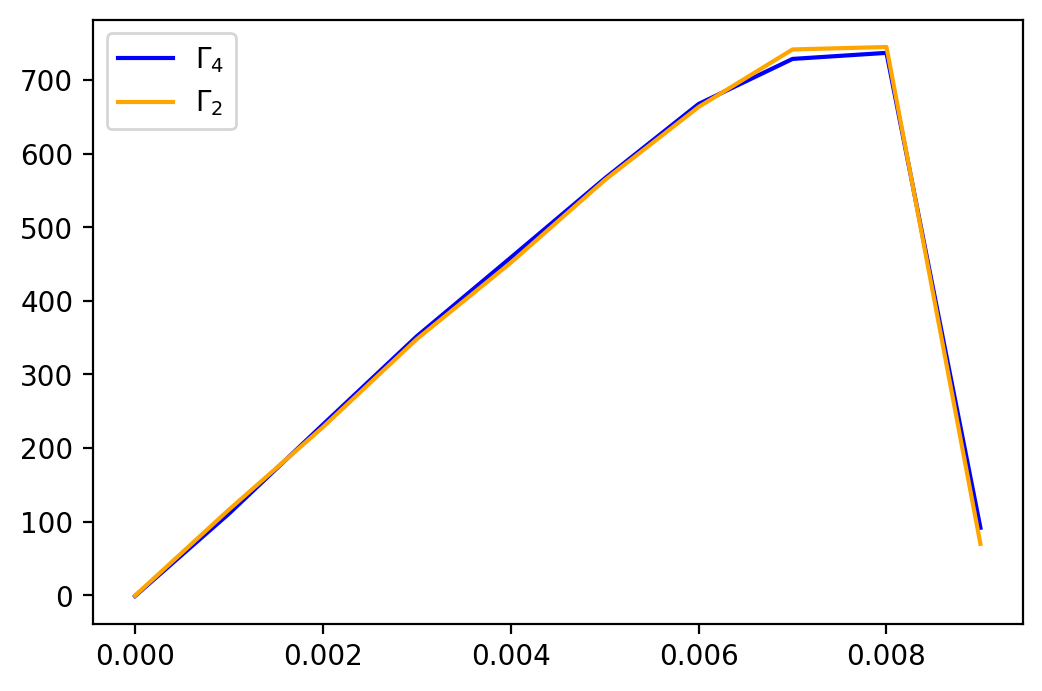}}\quad 
\subfloat[]{\includegraphics[width=1in,height=1in]{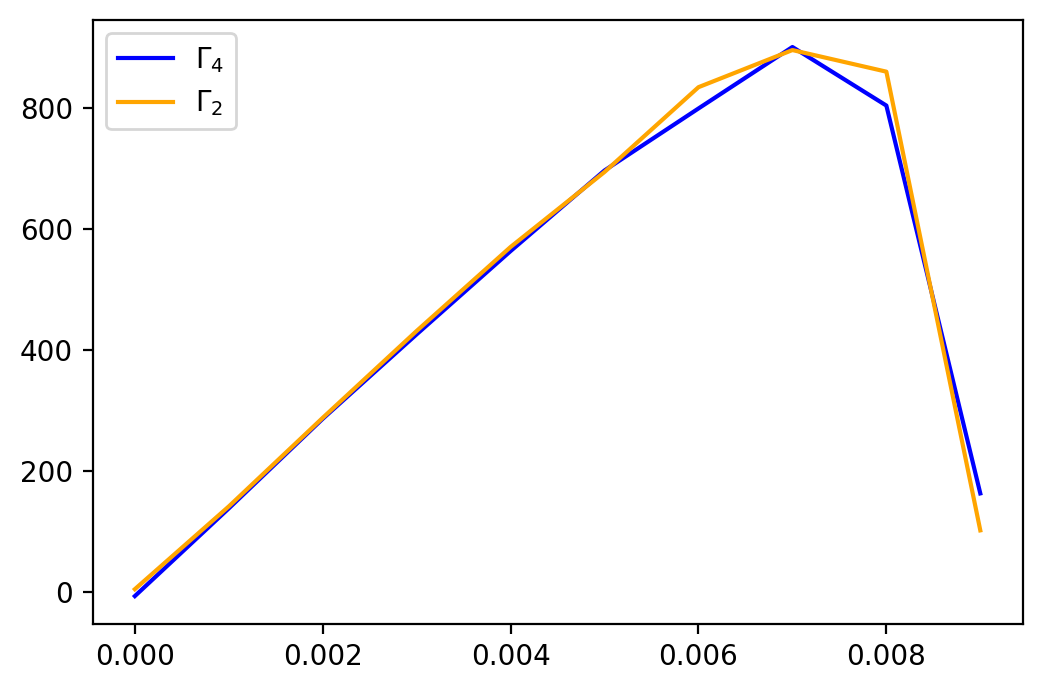}}
\caption{Force-displacement graph for second order and fourth order phase field model using the proposed approach for (a) $4$ subdomains, (b) $8$ subdomains, and (c) $12$ subdomains.}
\label{figp_ld}
\end{figure}

As the next example, we consider a unit square plate with a horizontal crack from the midpoint of the left outer edge to the center of the plate. The geometrical setup and the boundary conditions of the problem are shown in \autoref{fig:setup}(a). We consider $\lambda =  121.15$ kN/mm$^{2}$, $\mu =  80.77$ kN/mm$^{2}$ and $G_c = 2.7 \times 10^{-3}$ kN/mm. The computation is performed by applying a constant displacement increment of $\Delta u$ = $1\times 10^{-3}$ mm. In this example, we consider $l_0 = 0.0125$. 
The crack is initiated using the strain-history functional  as defined in \autoref{eq5}. The Dirichlet boundary conditions are:
\begin{equation}
    \boldsymbol u(0,y)=0, \;\;\boldsymbol v(x,0)=0, \;\; \boldsymbol v(x,1) = \Delta u,
\end{equation}
where $\boldsymbol u$ and $\boldsymbol v$ are the solutions of the elastic field in coordinate axes. In order to exactly satisfy the Dirichlet boundary conditions, the
neural network outputs for the elastic field are modified as:
\begin{equation}
\begin{split}
    \boldsymbol u&=[x(1-x)] \boldsymbol u_{\theta}, \\
    \boldsymbol v&=[y(y-1)]\boldsymbol v_{\theta}+y \Delta u, 
\end{split}
\end{equation}
where $\boldsymbol u_{\theta}$ and $\boldsymbol v_{\theta}$ are the neural network approximations of the displacement field in $x-$ and $y-$ axes. In this example an adaptive refinement scheme has been used. For robust implementations, subdomains within the cracked zones are refined and hence sub-divided into several sub-elements. We carried out experiments using multi subdomains, 4, 8 and 12 to understand the behaviour of the solution as the number of subdomains increases. \autoref{T1} gives the details of training points used in various experiments. The subdomain arrangement for all the experiments is shown in \autoref{fig:setup}(b)-(d). 
\begin{table}[htbp!]
    \caption{Summary of the subdomain arrangements studied in problem 2.}
    \label{T1}
    \begin{tabular}{cccc}
        \hline
        \textbf{\# of } & \textbf{\# elements} & \textbf{Integration} & \textbf{Interface points}\\
         \textbf{domains} & \textbf{per subdomain} & \textbf{points per element} & \textbf{per subdomain} \\ \hline
         $4$ & $11^2$ & $4$  & $1600$ \\
         $8$ &  $9^2$,$16^2$ & $4$  & $1000$ \\
         $12$ & $6^2$,$8^2$, $14^2$  & $4$  & $800$ \\
         \hline
    \end{tabular}
\end{table}
In all the cases, the networks has $4$- hidden layers with $50$ neurons in the model architecture, employing one DNN for each subdomain. As expected, the crack is less diffused for a smaller length scale parameter. The crack path obtained using $12$ subdomains and the fourth-order phase field model is shown in \autoref{figp4}. The presented crack patterns are the same as those reported in~\cite{goswami2019adaptive}. The cracks propagate in a horizontal direction. The load-displacement curves computed at the top edge of the plate are also shown in \autoref{figp_ld} with a comparison between fourth order and second order phase field models. Based on the plots, we can conclude that using the fourth-order model, we obtain a more linear response before failure. Also the post-failure behavior of the fourth-order model agrees well with the standard FEM and IGA \cite{goswami2020aadaptive}. Based on the plots we can also conclude that as the crack progresses, the refinement procedure refines its path locally. By increasing the density of the quadrature and interface points in each sub elements the imprinting effect of the interfaces has been significantly reduced. Moreover, we have used BFGS optimizer following SGD with learning rate $\in(0.001,0.09]$. Although large scaling factor speeds up the convergence rate but because of SGD optimizer one should avoid using larger scaling factor otherwise the model may encounter gradient explode or doesn't even converges at all. To overcome this difficulty we have added a regularization term so that it suppress the oscillations in the loss functions. In practice we  rely on trial and error analysis approach to learn a suitable value of regularization weight.

\subsection{Square plate with eccentric hole}

\begin{figure}[htbp]
\centering
\subfloat[]{\includegraphics[width=1.20in,height=1.20in]{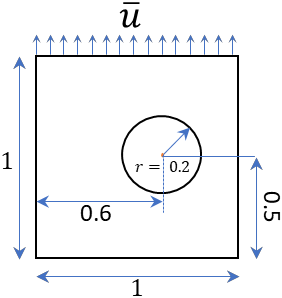}} \quad
\subfloat[]{\includegraphics[width=1.0in,height=1.0in]{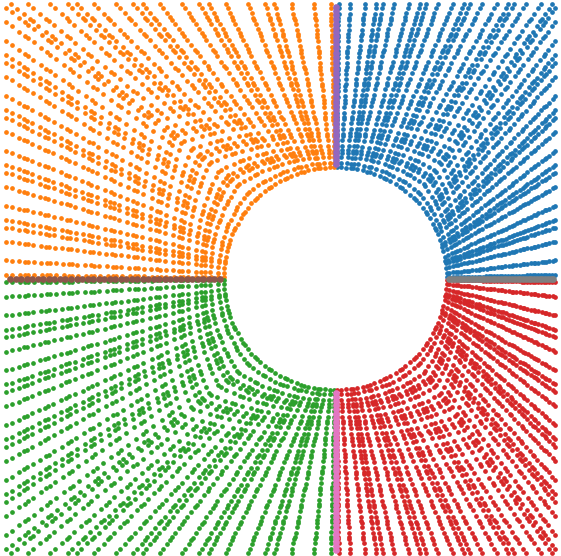}} \quad
\subfloat[]{\includegraphics[width=1.0in,height=1.0in]{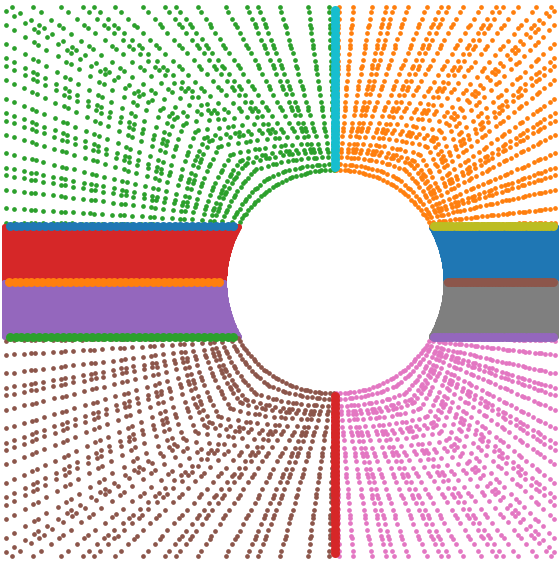}}
\caption{(a) Geometrical setup and boundary conditions for the square plate with eccentric hole test. All the units are in mm. (b) and (c) Distribution of training data over four and eight subdomains, respectively.}
\label{figp5}
\end{figure}

This numerical example is based on a unit square plate embedded with an eccentric hole, as shown in \autoref{figp5}(a). The bottom boundary remains fixed. At the top boundary, a quasi-static load is applied in the form of a prescribed displacement increment $\bar{\boldsymbol{u}}$. The material properties considered for this problem are: $\lambda= 121.154$ kN/mm$^2$, $\mu=80.77$ kN/mm$^2$, $G_c =  2.7 \times 10^3$  kN/mm. For this example, we have considered $l_0 = 0.02$. A displacement increment of $10^{-3}$ mm is applied at each incremental step.

\begin{figure}[!htbp]
\centering
\subfloat {\includegraphics[width=1.1in,height=0.9in]{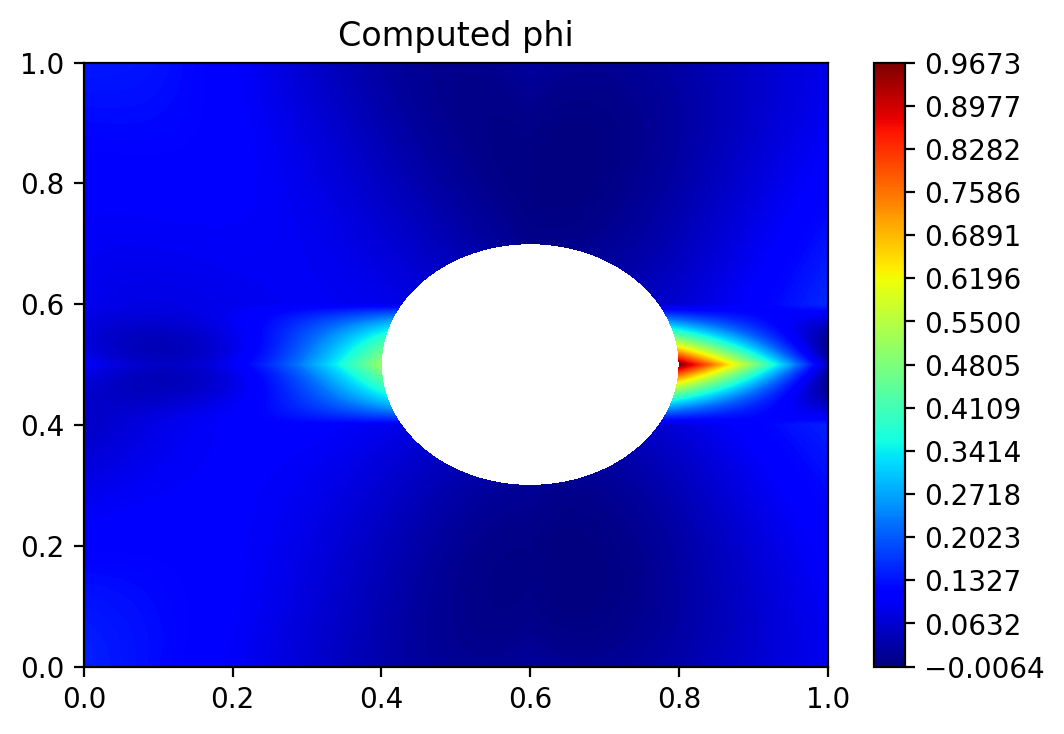}}\quad
\subfloat {\includegraphics[width=1.1in,height=0.9in]{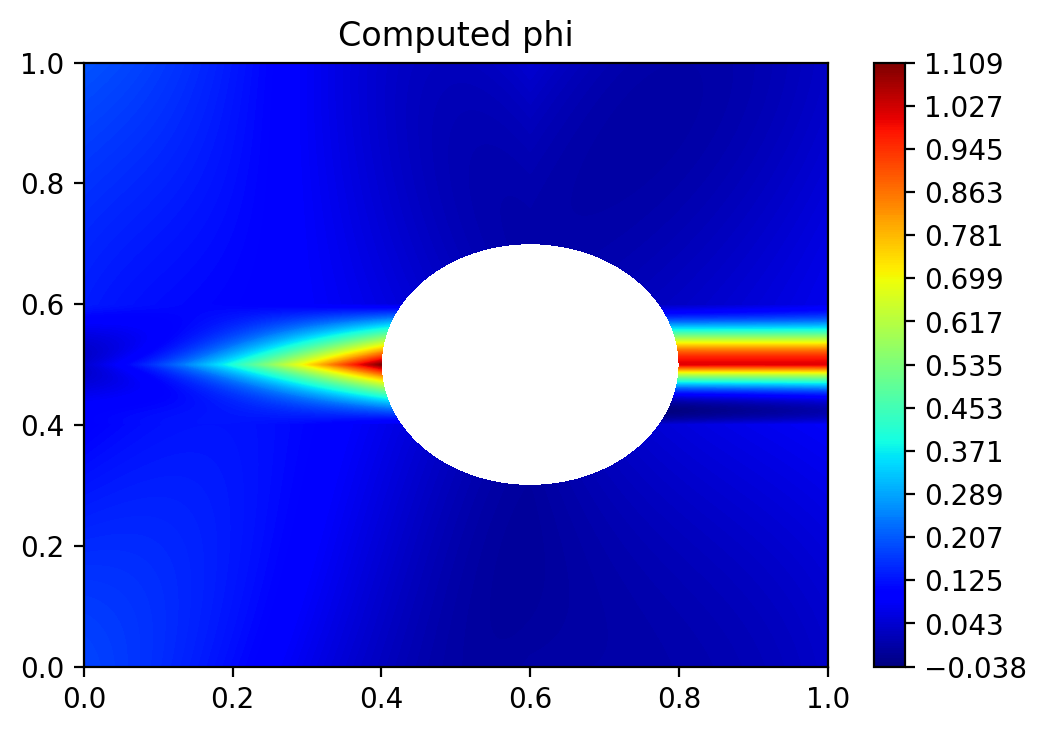}}\quad
\subfloat {\includegraphics[width=1in,height=0.9in]{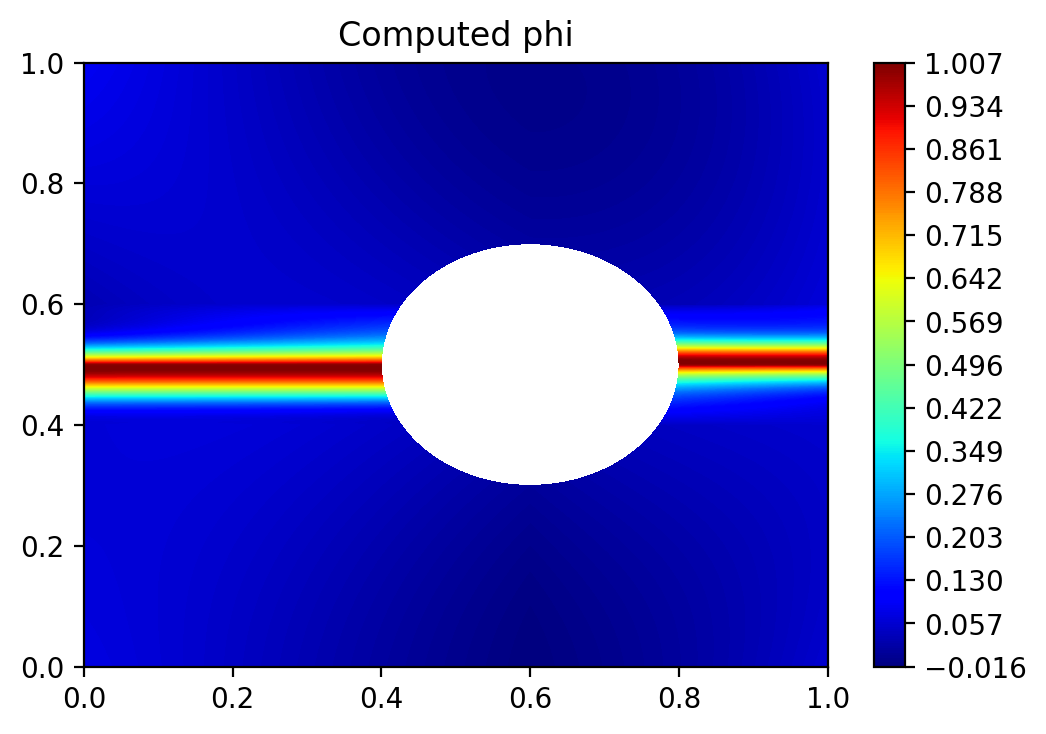}}
\caption{Crack growth in the square plate with eccentric hole, using the fourth-order phase field model and $8$ subdomains.}
\label{figp6}
\end{figure}

\begin{figure}[!htbp]
\centering
\subfloat[]{\includegraphics[width=1.5in,height=1.5in]{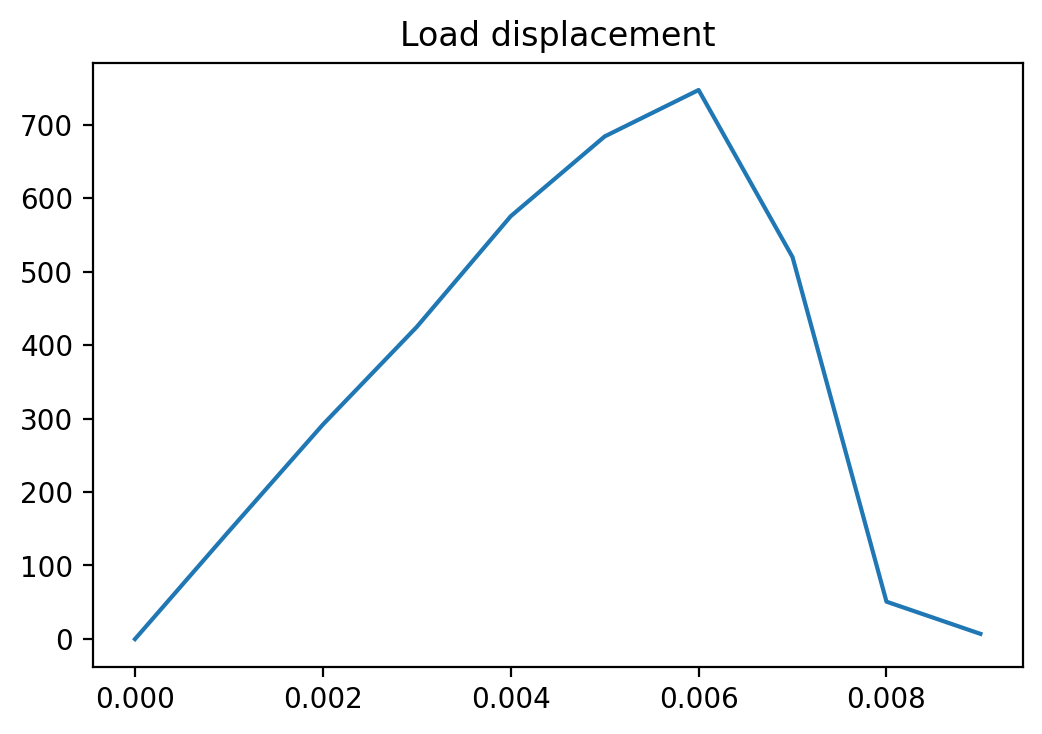}}\quad
\subfloat[]{\includegraphics[width=1.5in,height=1.5in]{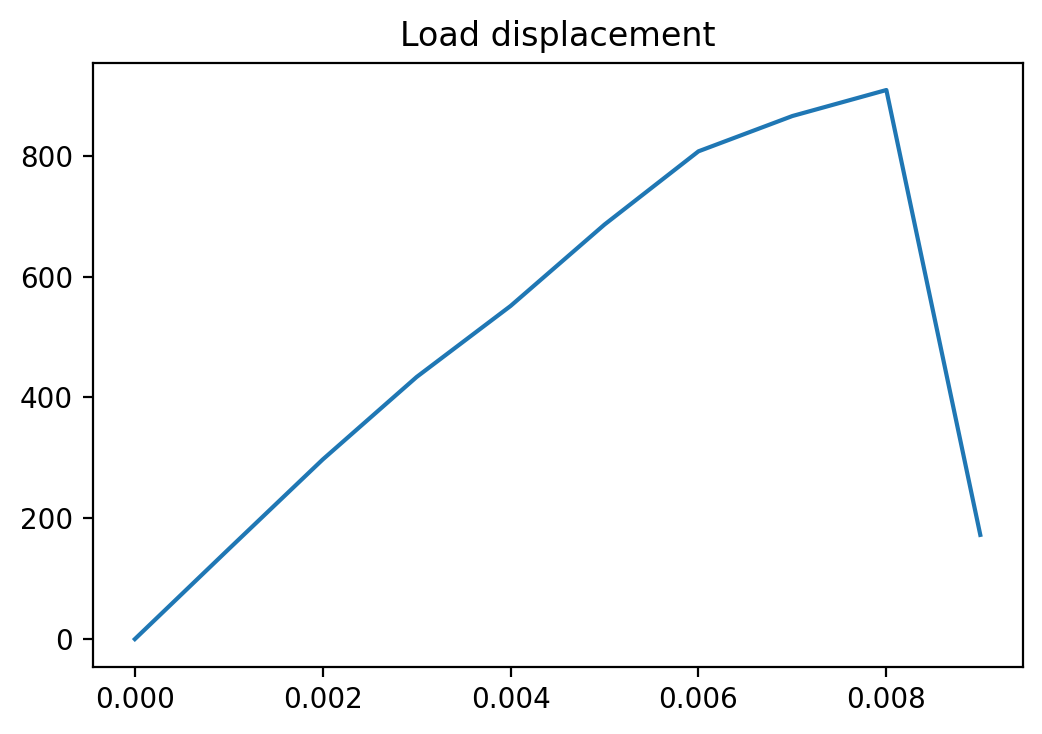}}
\caption{Load displacement plots of square plate with eccentric hole. (a) $4$ subdomains, and (b) $8$ subdomains.}
\label{figp7}
\end{figure}

\begin{table}[htbp!]
    \caption{Summary of the subdomain arrangements studied in problem 3.}
    \label{T2}
    \begin{tabular}{cccc}
        \hline
        \textbf{\# of } & \textbf{\# elements} & \textbf{Integration} & \textbf{Interface points}\\
         \textbf{domains} & \textbf{per subdomain} & \textbf{points per element} & \textbf{per subdomain} \\ \hline
         $4$ & $12^2$ & $4$  & $1600$ \\
         $8$ & $8^2$,$14^2$ & $4$  & $1000$ \\
         \hline
    \end{tabular}
\end{table}
In \autoref{T2}, the details of the experimental setup of $4$ and $8$ subdomains are provided. A $3$-layers fully-connected neural networks with $50$ neurons per hidden layer with $swish$ activation is employed. The final layer uses linear activation function. The outputs for the elastic field are altered to exactly match the Dirichlet boundary conditions, following:
\begin{equation}
\begin{split}
    \boldsymbol u& = x \boldsymbol u_{\theta},\\
    \boldsymbol v& = y(y-1) \boldsymbol v_{\theta}+ y \Delta u,
\end{split}
\end{equation}
where $\boldsymbol u$ and $\boldsymbol v$ are the solutions of the elastic field in coordinate axes, and $\boldsymbol u_{\theta}$ and $\boldsymbol v_{\theta}$ are the neural network approximations of the displacement field in $x-$ and $y-$ axes. \autoref{figp6} demonstrates the results of the fourth-order phase-field model obtained using the proposed adaptive XPINN based approach. The load-displacement curves for both the experiments are shown in \autoref{figp7}. In the specimen $8$ subdomains, we observe a linear pre-peak behavior while this linear stage is missing for the experiment with $4$ subdomains.

\section{Conclusion}
\label{sec:conclusion}
In this work, we have proposed a domain decomposition approach for variational physics informed neural networks (VE-XPINNs), that aims to accurately resolve the crack path employing the phase field modeling approach. The proposed approach is more efficient and accurate than its predecessor VE-PINNS \cite{goswami2020transfer, goswami2020adaptive}. The proposed method can be used to solve any differential equation in its weak form, much like the VE-PINN method can. This is accomplished by enforcing the residual continuity requirement along the shared subdomain interfaces. VE-XPINNs has the ability to be scaled up to solve real world probelms because of the deployment of separate neural network in each subdomain (which can be easily parallelized), and efficient and tailored hyper-parameter adjustment based on domain requirements. To speed up the learning ability of the networks and for faster convergence, we have employed adaptive activation functions for the mapping of the neural networks. The proposed approach is applied for solving three fracture mechanics examples. For all the examples, we observe that the results obtained using the proposed approach match closely with results from the literature.

\section*{Acknowledgment}
This work was supported in part by the Carl Zeiss Foundation (project ``Functionalisation of smart materials under multi-field requirements for transport infrastructure'') and ERC Starting Grant no: 802205.

\bibliographystyle{abbrv}  
\bibliography{references}
\end{document}